\documentclass[aps,prd,nofootinbib,nobibnotes,notitlepage,twocolumn]{revtex4-2}
\usepackage{graphicx}
\usepackage{amssymb}
\usepackage{amsmath} 
\usepackage{color}
\usepackage{float}
\usepackage{ulem}
\usepackage{accents}

\usepackage{dsfont}
\usepackage{bbm} 

\usepackage{amsfonts}
\usepackage[colorlinks=true,
pdfstartview=FitV,linkcolor=blue,
citecolor=blue,urlcolor=blue,breaklinks=true]{hyperref}
\usepackage{float}
\usepackage{placeins}
\usepackage[dvipsnames]{xcolor}
\usepackage{csquotes}
\usepackage[makeroom]{cancel}
\usepackage{units}
\usepackage{fixmath}
\usepackage{bigints}
\usepackage{amssymb}
\usepackage{systeme}
\renewcommand{\eqref}[1]{\mbox{Eq.~(\ref{#1})}}

\definecolor{ForestGreen}{rgb}{0.13,0.55,0.13}

\makeatletter
\renewcommand*\l@section{\@dottedtocline{1}{0em}{1.5em}}
\renewcommand*\l@subsection{\@dottedtocline{1}{1.5em}{1.5em}}
\renewcommand*\l@subsubsection{\@dottedtocline{1}{3em}{1.5em}}
\makeatother

\usepackage{capt-of}

\usepackage[caption=false]{subfig}
\begin{document}

\title{Kerr rotation signature of nonlinear Maxwell electrodynamics under a uniform electromagnetic background}

\author{M. J. Neves$^{a}$}
\email{mariojr@ufrrj.br}
\author{Pedro D. S. Silva$^{b}$}
\email{pedro.dss@ufma.br}
\affiliation{$^{a}$Departamento de F\'isica, Universidade Federal Rural do Rio de Janeiro, BR 465-07, CEP 23890-971, Serop\'edica, RJ, Brazil}
		\affiliation{$^{b}$Coordena\c{c}\~ao do Curso de Ci\^encias Naturais - F\'isica, Universidade Federal do Maranh\~ao, Campus de Bacabal, Bacabal, Maranh\~ao, 65700-000, Brazil}

\begin{abstract}

Nonlinear electrodynamics naturally arises in quantum field theory, where the electromagnetic vacuum behaves as an effective nonlinear optical medium, leading to phenomena such as vacuum birefringence and dichroism. Among the recently proposed models, modified Maxwell electrodynamics (ModMax) stands out as a conformally invariant nonlinear extension of Maxwell theory that preserves the fundamental symmetries of classical electrodynamics while predicting nontrivial optical effects. In this work, we investigate optical effects in ModMax electrodynamics in the presence of an external electromagnetic field. Considering uniform and constant magnetic and electric backgrounds, the solutions for the refractive indices are revisited. Using these results, we obtain the propagating modes and the phase shift (birefringence) for plane wave solutions in the presence of a pure magnetic background field. Afterwards, we investigate the Goos–Hänchen effect considering the interface between a simple dielectric and a medium whose electromagnetic response tensors are ruled by the ModMax electrodynamics. Further, based on the general reflection problem, we discuss the complex Kerr rotation with both the electric (E) and magnetic (B) background fields, considering two main cases: i) $B>E$ and ii) $E>B$. Our findings indicate that the $\gamma$ parameter and the ratios $(B/E)$ and $(E/B)$ play a central role in describing the Kerr signals (rotation and ellipticity) of systems with optical effects induced by nonlinear electromagnetic interactions.

\end{abstract}

\maketitle


%
%
%
\section{Introduction}
\label{sec:1}
Vacuum can be regarded as a nonlinear optical medium, a concept that naturally emerges in the Standard Model of elementary particles \cite{Battesti}. In this context, nonlinear electrodynamics models may exhibit phenomena such as vacuum dichroism and birefringence. In particular, vacuum magnetic birefringence (VMB) is a macroscopic quantum effect predicted by quantum electrodynamics (QED) \cite{EulerHeisenberg, Weisskopf}, in which the refractive indices for light polarized parallel and perpendicular to an external magnetic field differ in a nontrivial way.
The PVLAS (Polarisation of Vacuum with LASER) experiment devoted 25 years of continuous effort to the search for vacuum birefringence and dichroism. Although it has not yet reached the sensitivity required to observe the QED-predicted values, it has established the most stringent experimental bounds to date \cite{Zavattini,25years, DellaValle, Zavattini_Universe}. While VMB has not been directly detected in laboratory conditions, indirect evidence has been reported in the neutron star $\mbox{RX}\; \mbox{J}1856.5 - 3754$, which possesses magnetic fields on the order of $10^{13} \, \text{Gauss}$ \cite{Mignani}. 

Originally, the first non-linear electrodynamics (ED) was proposed by M. Born and L. Infeld to explain the static electric field of a pointlike charge particle at the origin \cite{Born}. Some years later, W. Heisenberg and H. Euler obtained the effective lagrangian for fermions interacting with an external electromagnetic field in QED, which leads to non-linear effects from quantum corrections at one loop \cite{EulerHeisenberg}. Thereby, these are the two most well-known non-linear EDs in the literature. After some decades, there are many others examples of non-linear EDs that includes several applications in research areas as: Superstring theories \cite{Fradkin,Pope}, Black hole solutions \cite{Banerjee,Mann}, Magnetic Monopoles \cite{Garcia}, Physics beyond the Standard Model \cite{Ellis}, Material physics \cite{Keser}, Axion physics \cite{Murchikova,PaixaoJHEP}, Plasma \cite{Haas0,Haas}, and others \cite{Plebanski,Boillat,Birula1,Birula2,Sorokin,TTdeformations1,TTdeformations2,NiauEPJC,NiauPRD,MJNevesEDN,MJNevesPat}. A new non-linear electrodynamics that preserves all the symmetries, including the conformal invariance, has been proposed and is known in the literature as modified Maxwell electrodynamics (ModMax) \cite{Bandos,Lechner}. The ModMax ED is also a possible solution for black holes \cite{Kruglov}, contains connection with supersymmetry \cite{BandosSusy}, Brane physics \cite{BandosBrane},  and exhibits the birefringence effect \cite{Mario-ModMax}. Since the ModMax is a relatively new non-linear ED, it motivates us to investigate other optical phenomena beyond the birefringence, such as the Goos-H\"anchen effect \cite{Goos}, polarization rotation, and the complex Kerr rotation.    

The reflection of light on surfaces constitutes a useful way for investigating the properties of distinct media, including the generation of circularly polarized and elliptically polarized waves, which are not commonly available in nature\cite{Miller}. These polarization states may emerge upon reflection from materials characterized by a complex refractive index (such as metals and chiral media) or under conditions of total internal reflection.

When an incident electromagnetic wave impinges on the surface of a gyrotropic medium, the reflected wave generally becomes elliptically polarized. The resulting polarization angles (rotation and ellipticity) define the Kerr rotation \cite{Sato, Argyres, Shinagawa},
 which is extensively employed to probe the optical and magneto-optical properties of materials. For instance, in Weyl semimetals \cite{Kargarian, Ghosh, Sekine}, where the axion term yields non-trivial optical responses (frequency-dependent Kerr signals) \cite{Trepanier}. These properties have been identified as relevant aspects in optical devices, such as chiral filters, circular polarizers, and optical isolators, since one can prevent the propagation of RCP and LCP waves within selected frequency intervals \cite{Cote-Trepanier, Cheng-Guo}.

In most material systems, however, Kerr rotations are typically small, amounting to less than 1 degree in conventional materials \cite{Schlenker-Souche} and lying in the range of $10^{-6}$ to $10^{-4}$ radians for topological insulators \cite{Sonowal}. The analysis of Kerr rotation has also served as an important diagnostic tool in the investigation of diverse physical systems, including precision measurements in quantum systems \cite{Tong-Li}, unconventional superconductors \cite{Kapitulnik}, and media described within CPT-even Lorentz-symmetry-violating electrodynamics \cite{Ruiz-Escobar}.

Most recently, optical signatures of axion dielectrics have been reported \cite{PRB-Pedro-Ronald}, where the chiral magnetic conductivity enables several effects, such as partial Brewster angles for polarization control, Goos-H\"anchen shift, giant Kerr signals, and conversion of a linearly polarized wave into a circularly polarized wave.

The large number of investigations addressing the electromagnetic properties of materials and nonconventional physical systems via optical phenomena, along with research on different non-linear electrodynamics theories, provides strong motivation to examine the optical effects that may emerge in systems governed by ModMax electrodynamics.

In this work, we investigate some optical signatures of the ModMax ED, such as the reflection properties based on the Goos-H\"anchen and in the complex Kerr effect. 
To do this, we consider the approach of the ModMax ED in an electromagnetic background that modifies the dispersion relations of a wave propagating through the material medium. Thus, we determine the contributions due to ModMax's non-linearity in relation to refractive indices and evaluate their influence on optical effects. These aspects can provide insights into new investigations as a nonlinear optical medium.

This paper is outlined as follows. In the section \ref{sec2}, we revisit the ModMax ED in the presence of an electromagnetic background field. In the section \ref{sec3},
we study the propagation modes (\ref{subsec3A}) and phase shift (\ref{subsec3B}) of the ModMax in a magnetic background field. The section \ref{sec4} is dedicated to the Goos-Hanch\"en effect in ModMax ED in the presence of an external magnetic field. Afterwards, in the section \ref{sec5}, we investigate the complex Kerr rotation when both the electric $(E)$ and magnetic $(B)$ backgrounds are present under the conditions of $B > E$ and $E > B$, respectively. Finally, we summarize our results in section \ref{sec6}.

We use the natural units $\hbar=c=1$  with $4 \pi \epsilon_0 = 1$, and the Minkowski
metric signature adopted is $\eta^{\mu\nu}=\mbox{diag}(+1,-1,-1,-1)$. In natural units, the electric and magnetic fields have squared-energy mass dimension where 
the conversion of Volt/m and Tesla (T) to the natural system is as follows: $1 \, \mbox{Volt/m}=2.27 \times 10^{-24} \, \mbox{GeV}^2$ and $1 \, \mbox{T} =  6.8 \times 10^{-16} \, \mbox{GeV}^2$, respectively.

\section{Basics on ModMax ED in an electromagnetic background field}
\label{sec2}

In this section, we review some results obtained in the ref. \cite{Mario-ModMax}.
The non-linear ModMax ED is governed by the Lagrangian density

\begin{eqnarray}\label{ModMaxL}
{\cal L}_{MM}({\cal F}_{0},{\cal G}_{0})=\cosh\gamma \, {\cal F}_0 + \sinh\gamma \, \sqrt{{\cal F}_0^2+{\cal G}_0^2} \; ,
\end{eqnarray}
where $\gamma$ is a real parameter that satisfies the condition $\gamma \geq 0$, and it depends on the Lorentz- and gauge-invariant bilinears
$
{\cal F}_{0}=-\frac{1}{4} \, F_{0\mu\nu}^{2}=\frac{1}{2} \, \left( {\bf E}_{0}^2-{\bf B}_{0}^2\right) ,
$
and
$
{\cal G}_{0}=-\frac{1}{4} \, F_{0\mu\nu}\widetilde{F}_{0}^{\;\,\,\mu\nu}={\bf E}_{0}\cdot{\bf B}_{0}.
$

The electromagnetic field strength tensor is $F_{0}^{\;\,\,\mu\nu}=\partial^{\mu}A_{0}^{\;\,\nu}-\partial^{\nu}A_{0}^{\;\,\mu}=\left( \, -E_{0}^{\;\,i} \, , \, -\epsilon^{ijk}B_{0}^{\;\,k} \, \right)$, whose dual tensor is $\widetilde{F}_{0}^{\;\,\,\mu\nu}=\epsilon^{\mu\nu\alpha\beta}F_{0\alpha\beta}/2=\left( \, -B_{0}^{\;\,i} \, , \, \epsilon^{ijk}E_{0}^{\;\,k} \, \right)$, which satisfies the equations without sources, {\it i.e.}, $\partial_{\mu}\widetilde{F}_{0}^{\;\,\mu\nu}=0$.

The electromagnetic background field is so introduced by the expansion of the $4$-potential as 
$A_{0}^{\;\,\mu}(x)=a^{\mu}(x)+A_{B}^{\;\;\,\,\mu}(x)$, where $a^{\mu}$ is the photon propagating field,
and $A_{B}^{\;\;\,\,\mu}$ is the background $4$-potential, that in case of an uniform and constant background field satisfies the relation  $A_{B}^{\;\;\,\,\mu}=-F_{B\mu\nu}\,x^{\nu}/2$. As consequence, the strength field tensor $F_{0}^{\;\,\,\mu\nu}$ is written
as $F_{0}^{\;\,\,\mu\nu}=f^{\mu\nu} \, + \, F_{B}^{\;\;\,\mu\nu}$, where 
$f^{\mu\nu}=\partial^{\mu}a^{\nu}-\partial^{\nu}a^{\mu}=\left( \, -e^{i} \, , \, -\epsilon^{ijk}b^{k} \, \right)$ 
is the propagating electromagnetic field in the space-time, and $F_{B}^{\;\;\,\mu\nu}=\left( \, -E^{i} \, , \, -\epsilon^{ijk}B^{k} \, \right)$ is the 
electromagnetic background field, which we consider uniform and constant throughout this paper. For convenience, the index $(B)$ in the tensors is related 
with the EM background field.

Since we are interested in propagation effects, the Lagrangian (\ref{ModMaxL}) expanded up to second order in the propagating tensor $f^{\mu\nu}$ is
\begin{align} 
\label{L4}
 {\cal L}_{MM}^{(2)}  &=  -\frac{1}{4} \, c_{1} \, f_{\mu\nu}^{\, 2}
-\frac{1}{4} \, c_{2} \, f_{\mu\nu}\widetilde{f}^{\mu\nu}
+\frac{1}{8} \, Q_{B\mu\nu\kappa\lambda} \, f^{\mu\nu}f^{\kappa\lambda} 
\; ,
\end{align}
where the background tensor $Q_{B\mu\nu\kappa\lambda}$ is defined by
\begin{align}
Q_{B\mu\nu\kappa\lambda} &= d_{1} \, F_{B \mu\nu} \, F_{B\kappa\lambda}
+d_{2} \, \widetilde{F}_{B \mu\nu} \, \widetilde{F}_{B \kappa\lambda}  
\nonumber \\
&\phantom{=}+d_{3} \, F_{B \mu\nu} \, \widetilde{F}_{B \kappa\lambda}
+ d_{3} \, \widetilde{F}_{B \mu\nu} \, F_{B \kappa\lambda} \; ,
\end{align}
and $\widetilde{F}_{B}^{\;\,\,\mu\nu}=\epsilon^{\mu\nu\alpha\beta}F_{B\alpha\beta}/2=\left( \, -B^{i} \, , \, \epsilon^{ijk}E^{k} \, \right)$ is the dual tensor of $F_{B\mu\nu}$. The coefficients $c_{1}$, $c_{2}$, $d_{1}$, $d_{2}$ and $d_{3}$ for the expansion from (\ref{L4}) are evaluated at the background fields 
${\bf E}$ and ${\bf B}$ as:

\begin{eqnarray}
c_{1}&=& \left.\frac{\partial{\cal L}_{MM}}{\partial{\cal F}_{0}}\right|_{{\bf E},{\bf B}}
, \,
c_{2}= \left. \frac{\partial{\cal L}_{MM}}{\partial{\cal G}_{0}}\right|_{{\bf E},{\bf B}}
,  \,
\left. d_{1}=\frac{\partial^2{\cal L}_{MM}}{\partial{\cal F}_{0}^2}\right|_{{\bf E},{\bf B}}
, \,  \label{coefficients} 
\nonumber \\
d_{2}&=& \left.\frac{\partial^2{\cal L}_{MM}}{\partial{\cal G}_{0}^2}\right|_{{\bf E},{\bf B}}
,   \,
d_{3}= \left. \frac{\partial^2{\cal L}_{MM}}{\partial{\cal F}_{0}\partial{\cal G}_{0}}\right|_{{\bf E},{\bf B}} . \label{coefficients-2}
\hspace{0.4cm}
\end{eqnarray}

These coefficients depend only on the EM field magnitude, and also depend on the ModMax $\gamma$-parameter. 
Using the lagrangian density (\ref{ModMaxL}), the coefficients are given by

\begin{eqnarray}\label{coefficientsMM}
c_{1} &=& \cosh\gamma+\frac{\sinh\gamma\,{\cal F}}{\sqrt{{\cal F}^2+{\cal G}^2}}
\; , \;
c_{2}=\frac{\sinh\gamma\,{\cal G}}{\sqrt{{\cal F}^2+{\cal G}^2}} \; ,
\nonumber \\
d_{1} &=& \frac{\sinh\gamma \, {\cal G}^{2}}{({\cal F}+{\cal G}^2)^{3/2}} 
\; , \;
d_{2} = \frac{\sinh\gamma \, {\cal F}^{2}}{ ({\cal F}^2+{\cal G}^2)^{3/2} } \; ,
\nonumber \\
d_{3} &=& \frac{\sinh\gamma \, {\cal F} \, {\cal G}}{ ({\cal F}^2+{\cal G}^2)^{3/2} } \; ,
\end{eqnarray}
in which ${\cal F}=({\bf E}^2-{\bf B}^{2})/2$ and ${\cal G}={\bf E} \cdot{\bf B}$ are the gauge and Lorentz
invariants related to ${\bf E}$ and ${\bf B}$. Notice that in the particular case where the electric background is perpendicular to the magnetic one, 
${\cal G}={\bf E}\cdot{\bf B}=0$, the coefficients are simplified as: $c_{2}=d_{1}=d_{3}=0$, and
\begin{eqnarray}\label{c1d2EperpB}
c_{1} = \cosh\gamma+\mbox{sgn}({\cal F}) \, \sinh\gamma 
\; \; , \; \;
d_{2} = \sinh\gamma \, \frac{{\cal F}^{2}}{ |{\cal F}|^{3} } \; ,
\end{eqnarray}
where $\mbox{sgn}({\cal F})$ is the sign function of ${\cal F}$, with $\mbox{sgn}({\cal F})=+1$, if ${\cal F}>0$, 
and $\mbox{sgn}({\cal F})=-1$, if ${\cal F} < 0$. The expressions (\ref{c1d2EperpB}) is the hardest case that leads 
in the paper for our analysis of optics effects. 
Under these conditions, the field equations for the Lagrangian (\ref{L4}) are read as  
\begin{subequations}
\begin{align}
\label{EqGmunu}
 \partial^\mu \left[ \, c_1 \, f_{\mu \nu} -
\frac{d_2}{2} \, \widetilde{F}_{B \mu\nu} \left(\widetilde{F}_{B \kappa\lambda} f^{\kappa \lambda}\right) \, \right]=0 
\; ,
\\
\partial_{\mu}\widetilde{f}^{\mu\nu}=0 \; .
\end{align}
\end{subequations}
In the vector form, the previous equations are
\begin{subequations}
\begin{eqnarray}
\nabla\cdot{\bf D}&=&0 \;\;\; , \;\;\;
\nabla\times{\bf e}+\frac{\partial {\bf b}}{\partial t} = {\bf 0} \; ,
\label{divDrote}
\\
\nabla\cdot{\bf b}&=&0 \;\;\; , \;\;\;
\nabla\times{\bf H}-\frac{\partial {\bf D}}{\partial t} = {\bf 0} \; ,
\label{divbrotH}
\end{eqnarray}
\end{subequations}
where the vectors ${\bf D}$ and ${\bf H}$ have the components
\begin{subequations}
\label{general-nonlinear-constitutive-relations-1}
\begin{align}
D_{i} &= \epsilon_{ij} \, e_{j}+\sigma_{ij} \, b_{j} \; ,
\label{D0i}
\\
H_{i} &= -\sigma_{ji} \, e_{j}+ (\mu^{-1})_{ij} \, b_{j}
\; ,
\label{H0i}
\end{align}
\end{subequations}
that define the electric permittivity tensor $\epsilon_{ij}$,
$\sigma_{ij}$, and the inverse of the magnetic permeability tensor
\begin{subequations}
\begin{eqnarray} \label{eij}
\epsilon_{ij} &=& \delta_{ij} 
+ d_E \, B_{i} \, B_{j}
\; ,
\\
\sigma_{ij} &=& 
d_{E} \, B_{i} \, E_{j}
\; ,
\\
(\mu^{-1})_{ij} &=& \delta_{ij} 
- d_E \, E_{i} \, E_{j} 
\label{muijinv}
\; .
\end{eqnarray}
\end{subequations}
Here, for simplicity, we have defined $d_{E}=d_2/c_1$. The usual Maxwell ED is recovered when $\gamma \rightarrow 0$. 
%
%

%
%

%
\label{sec3}
Using the plane wave solutions for ${\bf e}$ and ${\bf b}$, the equations (\ref{divDrote})-(\ref{divbrotH}) 
yield the wave equation for the electric amplitude ${\bf e}_{0}$ :
\begin{eqnarray}\label{Mijej}
M^{ij} \, {\bf e}_{0}^{\,\,j}=0 \; ,
\end{eqnarray}
where the matrix elements $M^{ij}$ are given by
\begin{eqnarray}
\label{Mij}
M^{ij} &=& a \, \delta^{ij} + b \, n^i \, n^j
+d_{E} \, B^i \, B^j
\nonumber \\
&&
\hspace{-0.5cm}
-d_{E} \, {\bf n}^2 \, E^i \, E^j + \, d_{E} \left({\bf E\cdot n}\right)\left(E^i \, n^j + E^j \, n^i\right) 
\nonumber \\
&&
\hspace{-0.5cm}
+ d_{E} \left[ \, B^i \left( {\bf E} \times {\bf n} \right)^j + B^{j} \left( {\bf E} \times {\bf n} \right)^i \, \right]
\; ,
\end{eqnarray}
with $a=1-{\bf n}^2+d_{E} \left({\bf n} \times {\bf E}\right)^2$ and $b=1-d_{E} \, {\bf E}^2$, and we have defined the refractive 
index as ${\bf n}={\bf k}/\omega$, for a wave vector ${\bf k}$ and wave frequency $\omega$.
The non-trivial solution of (\ref{Mijej}) is when $\mbox{det}(M_{ij})=0$. Thus, this polynomial equation provides three solutions: $n_{1}=1$  
that confirms the gauge invariance of the model, $n_3=1$ in which we have $d_{B}=0$ for electric and magnetic backgrounds perpendiculars, 
and the others solutions that we are interested in this paper are given by  
\begin{widetext}
\begin{eqnarray}\label{nomegap}
n_{2EB}^{(\pm)}=
\frac{1+d_{E}\,{\bf B}^2}{ d_{E} \, {\bf B}\cdot(\hat{{\bf k}}\times{\bf E})\pm\sqrt{1-d_{E}(\hat{{\bf k}}\times{\bf E})^{2}
- d_{E}^2\left({\bf E}^2-{\bf B}^{2} \right)(\hat{{\bf k}}\cdot{\bf B})^2 + d_{E}\,(\hat{{\bf k}}\cdot{\bf B})^{2}+d_{E}\,{\bf B}^2 } } \; .
\end{eqnarray}
\end{widetext}
These last solutions provide the non-linear effects that we are interested for investigations through the Optics effects, 
as the phase shift, the Goos-H\"anchen and the complex Kerr rotation.

\section{Propagating modes and phase shift with magnetic background}
\label{sec3}

For the analysis of the propagating modes and phase shift, we consider only the magnetic background field, 
that is, in the solutions of the previous section, we make ${\bf E}={\bf 0}$. In this case, the coefficients $c_{1}$ and $d_{2}$ 
are 
\label{coefficientsMM}
\begin{align}
c_{1}=e^{-\gamma} 
\; \; \mbox{and} \; \;
d_{2} =  2\,\frac{\sinh\gamma}{{\bf B}^2} \; ,  
\label{modMax-coefficients-1} 
\end{align}
in which $d_{E}$ is
\begin{eqnarray}
d_{E}=\frac{d_2}{c_1}=2\,e^{\gamma} \, \frac{\sinh\gamma}{{\bf B}^2}=\frac{e^{2\gamma}-1}{{\bf B}^2} \; .
\end{eqnarray}
The non-trivial solutions of the refractive index in (\ref{nomegap}) reduces to : 
\begin{align}\label{n2B}
n_{2B}^2=\frac{1+d_{E}\,{\bf B}^{2}}{1 +d_{E} \,( {\bf{B}} \cdot  \hat{\bf{k}})^{2} } = \frac{e^{2\gamma}}{\sin^2\theta + e^{2\gamma} \cos^{2}\theta} \; ,
\end{align}
that does not depend on the magnetic field magnitude, but it depends on the $\theta$-angle that ${\bf B}$ does with the wave propagation direction ${\bf k}$, 
in which $\cos\theta=\hat{{\bf k}}\cdot\hat{{\bf B}}$. The refractive index $n_{1B}$ corresponds to the standard mode. On the other hand, the indices $n_{2B}$ include all the anisotropic contributions and the ModMax $\gamma$-parameter.
\begin{figure}[h]
\begin{centering}
\includegraphics[scale=0.6]{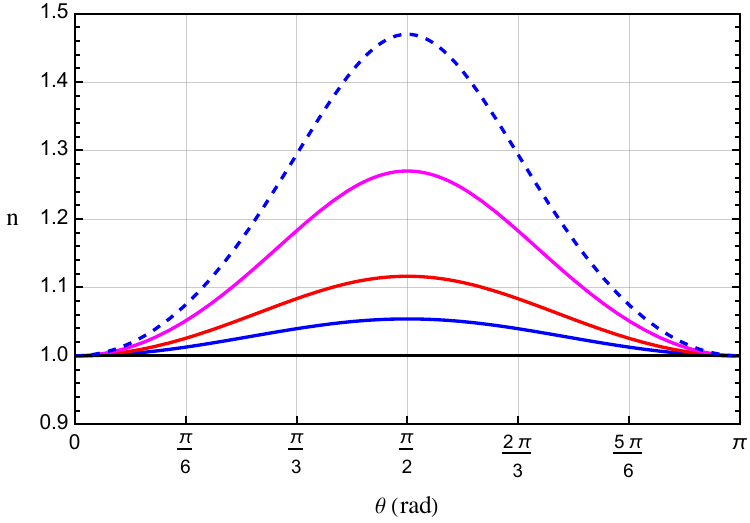}
\par\end{centering}
\caption{\small{\label{plot-indice-2B-plus-High}}Refractive index $n_{2B}$ of \eqref{n2B} in terms of propagation direction $\theta$.  Here, we have set: $\gamma=$ 0.1 (blue), 0.2 (red), 0.4 (magenta), 0.6 (dashed blue). The solid black curve represents the standard case with $\gamma \rightarrow 0$.}
\end{figure}
As indicated in the Fig.~\ref{plot-indice-2B-plus-High}, the maximum values of refractive index $n_{2B}$ happen when the direction of propagation is perpendicular to the background magnetic field, which means that birefringence effects can be enhanced when the magnetic field is perpendicular to the direction of light in the system. Furthermore, for propagation directions aligned to the magnetic field, that is, $\theta=\{0,\pi\}$, one finds $n_{2B} \rightarrow 1$, recovering the usual case. 
We apply this result in the propagating modes ahead.

%
%

%

\subsection{Propagating modes}
\label{subsec3A}

To determine the polarizations of the propagating modes, let us choose a coordinate system in which ${\bf B}= (0, 0, B)$ and ${\bf{n}}=n\,(\sin\theta, 0, \cos\theta)$. 
Thereby, by considering particular these cases for propagation direction and implementing the previous solutions for the refractive indices, $n_{1B}$ and $n_{2B}$, in \eqref{Mijej}, one finds that :
\begin{itemize}

\item[i)] For oblique directions of propagation with ${\bf{n}}=n\,(\sin\theta, 0, \cos\theta)$, with $\theta\neq \{0, \pi/2, \pi \}$, the propagating modes are
\begin{align}
n_{1B} &= 1 \; 
, \quad {\bf{e}}_{1B} =  \begin{pmatrix}
0 \\
1\\
0
\end{pmatrix} \; ,  \label{eq-nonlinear-1}
\end{align}
and
\begin{subequations}
\begin{eqnarray}\label{eq-nonlinear-2} 
n^{2}_{2B} &=& \frac{e^{2\gamma}}{\sin^2\theta + e^{2\gamma} \cos^{2}\theta} \; , \; 
\\
{\bf e}_{2B} &=& \frac{1}{\sqrt{1+f^{2}}} 
\left(
  \begin{array}{c}
    1 \\
    0 \\
    -f \\
  \end{array}
\right) \; ,  \;\;\;\;
\label{eq-nonlinear-3}
\end{eqnarray}
\end{subequations}
where $f = e^{-2\gamma}\, \tan \theta$, and $\theta$ is the angle between $\hat{{\bf k}}$ and ${\bf B}$.

\item[ii)] In Faraday configuration, ${\bf{n}}=(0, 0, n)$, $(\theta=0, \pi)$, it holds
\begin{align}
n_{1B}&= 1 \; , \quad 
 {\bf{e}}_{1B} = \frac{1}{\sqrt{1+ |a|^{2} } } \begin{pmatrix}
1 \\
a\\
0
\end{pmatrix} ,  \label{eq-nonlinear-5} \\
n_{2B} &=1 \; , \quad {\bf{e}}_{2B} = \frac{1}{\sqrt{1 + |a|^{2}}} \begin{pmatrix}
1\\
b\\
0
\end{pmatrix} \; ,  \label{eq-nonlinear-6}
\end{align}
with $a$ and $b$ being an arbitrary constants.
\item[iii)] For Voigt configuration $(\theta=\pi/2)$, one obtains
\begin{align}
 n_{1B}&= 1  \; , \;  \quad {\bf e}_{1B} =  \begin{pmatrix}
0 \\
1\\
0
\end{pmatrix} ,  \label{eq-nonlinear-9}
\end{align}
\begin{align}
n_{2B} &= e^{\gamma}  
\; \; , \; \; 
{\bf e}_{2B} = \begin{pmatrix}
0 \\
0 \\
1
\end{pmatrix} \; . 
\label{eq-nonlinear-10}
\end{align}

\end{itemize}

\subsection{Phase shift}
\label{subsec3B}
Let us consider the Faraday configuration given in item $ii)$ of Sec.~\ref{subsec3A}.  In this case, the propagating modes related to $n_{1B}$ and $n_{2B+}$ are the same since $n_{1B}=n_{2B+}$, for which the polarization vector is given generically by the last relation in \eqref{eq-nonlinear-1}. On the other hand, for the propagating mode associated with $n_{2B-}$, one finds a linearly polarized vector. Thereby, the birefringence is evaluated in terms of the phase shift \cite{Pedro-2021}, defined by
\begin{align}
\Delta =\frac{2\pi}{\lambda} \, d \, \Delta n \; ,  \label{eq-nonlinear-15}
\end{align}
where $\lambda$ is the wavelength of the electromagnetic radiation in the vacuum, $d$ corresponds to the thickness of the medium (or the distance the wave propagates through the medium), and $\Delta n$ is the difference between the two refractive indices, that in this case is $\Delta n=n_{2B}-n_{1B}$. Therefore, one finds

Considering the Voigt configuration, we obtain the phase shift
\begin{align}
\Delta = \frac{2\pi}{\lambda} \, d \, \left[ \, \sqrt{ 1 + 2\,e^{\gamma}\sinh\gamma } - 1 \, \right] \simeq \frac{2\pi}{\lambda} \, d \, \gamma  \; , \label{eq-nonlinear-17}
\end{align}
for a small $\gamma$-parameter.

\section{Goos-H\"anchen effect}
\label{sec4}
We consider the interface between a simple dielectric, which we call medium 1, with electric permittivity $\epsilon_{1}$ and magnetic permeability $\mu_{1}$, 
and medium 2 ruled by the ModMax electrodynamics. Considering the incident light coming from medium 1, and impinging on the surface of medium 2 with 
incidence angle $\theta_{I}$ greater than the critical angle $\theta_{c}$, given by
\begin{eqnarray}\label{eq-nonlinear-18}
\theta_{c} =\arcsin \left( \frac{n_{2B}}{\sqrt{\mu_{1}\epsilon_{1}}} \right) \; . 
\end{eqnarray}
where $n_{2B}$ is the solution (\ref{n2B}). The corresponding reflected wave can undergo a lateral shift, that is, a displacement of the reflected wave relative to the position in which there is no total reflection, defined by
\begin{eqnarray}\label{DGoos}
D = -\frac{\lambda}{2\pi} \frac{\partial \varphi}{\partial \theta_{I}} \; , \label{eq-nonlinear-19}
\end{eqnarray}
where $\lambda$ is the wavelength of the incident light, and $\varphi$ is the phase acquired during the total reflection. This lateral optical shift of the reflected ray is known as the Goos-H\"anchen shift \cite{Ming-Che, PRB-Pedro-Ronald, Jackson}. We explore this effect in the situations in which the incident wave is $s$- and $p$-polarized. 

\subsubsection{Incident wave $s$-polarized}
For an incident wave $s$-polarized, one can use $\mu_{1}=\mu_{2}=1$, where the medium ruled by the ModMax electrodynamics in the magnetic background case, we have 
$\mu_{2}=1$, according to \eqref{muijinv}, and $n_{1}=\sqrt{\mu_{1}\epsilon_{1}}$ for the linear dielectric medium, which yields the following Fresnel coefficient
\begin{align}
r^{s}_{2B}&= \frac{ \cos\theta_{I} -  \sqrt{ n_{2B}^{2}/\epsilon_{1} - \sin^{2}\theta_{I}}}{  \cos\theta_{I} +  \sqrt{ n_{2B}^{2}/\epsilon_{1}  - \sin^{2}\theta_{I}}} \; . \label{eq-nonlinear-21}
\end{align}
In the total reflection regime, it holds $\sin \theta_{I} > \sin \theta_{c}$, then \eqref{eq-nonlinear-21} becomes
\begin{align}\label{eq-nonlinear-22}
r^{s}_{2B}= e^{i\varphi^{s}_{2B}} \; , 
\end{align}
where the phase is given by
\begin{align}\label{eq-nonlinear-23}
\varphi^{s}_{2B} = -2 \arctan \left[ \, \sec\theta_{I} \, \sqrt{\sin^{2}\theta_{I}- \frac{n_{2B}^{2}}{\epsilon_{1}}  } \, \right] \; .  
\end{align}
Substituting (\ref{eq-nonlinear-23}) in the definition (\ref{DGoos}), the Goos-H\"anchen effect for $s$-wave polarized is
\begin{eqnarray}
\label{Ds}
D_{s}=\frac{\lambda}{\pi}\,\sin\theta_{I} \left[ \, \sin^2\theta_{I}-\frac{n_{2B}^2}{\epsilon_{1}} \, \right]^{-1/2} \; .
\end{eqnarray}
Notice that, in general, the result (\ref{Ds}) is an implicit function of the $\gamma$-ModMax parameter and of the $\theta$-angle of 
$\cos\theta=\hat{{\bf k}}\cdot\hat{{\bf B}}$ due to $n_{2B}$. For the Faraday configuration, we have $n_{2B}=1$, 
the expression (\ref{Ds}) is reduced to the simplest case :
%
%
\begin{eqnarray}\label{DsFaraday}
D_{s}=\frac{\lambda}{\pi}\,\sin\theta_{I} \left[ \, \sin^2\theta_{I}-\frac{1}{\epsilon_{1}} \, \right]^{-1/2} \; .
\end{eqnarray}
%
%
%
In the Voigt scenario, we obtain
%
%
%
\begin{eqnarray}\label{DsVoigt}
D_{s}(\gamma)=\frac{\lambda}{\pi}\,\sin\theta_{I} \left[ \, \sin^2\theta_{I}-\frac{e^{2\gamma}}{\epsilon_{1}} \, \right]^{-1/2} . \; \; \;
\end{eqnarray}
%

%
%
%

%
\subsubsection{Incident wave $p$-polarized}
For an incident wave $p$-polarized in the total reflection regime, we obtain
\begin{align}
r^{p}_{2B} &=e^{i\varphi^{p}_{2B}} \; ,  \label{eq-nonlinear-29}
\end{align}
with the reflection phase given by
\begin{align}
\varphi^{p}_{2B}&= - 2 \arctan \left[ \, \frac{\epsilon_{1}}{n_{2B}^{2} \cos\theta_{I}} \, \sqrt{\sin^{2} \theta_{I} - \frac{n_{2B}^{2}}{\epsilon_{1}} } \, \right] \; , \label{eq-nonlinear-30}
\end{align}
and using (\ref{DGoos}), we obtain
\begin{eqnarray}\label{Dp}
D_{p} &=& \frac{\lambda \, \epsilon_{1}}{\pi} \frac{n_{2B}^2 \, \sin\theta_{I} }{n_{2B}^4\cos^2\theta_{I}-n_{2B}^2\,\epsilon_{1}+\epsilon_{1}^2 \, \sin^2\theta_{I} } 
\nonumber \\
&&
\times \, \frac{1-n_{2B}^{2}/\epsilon_{1}}{ \sqrt{ \sin^2\theta_{I}-n_{2B}^2/\epsilon_{1} } } \; .
\end{eqnarray}
In the Faraday configuration, we have
\begin{equation}\label{DpFaraday}
D_{p} =
\frac{\lambda}{\pi} \frac{ \sin\theta_{I} }{1-\epsilon_{1}+(\epsilon_{1}^2-1) \sin^2\theta_{I} } 
\frac{\epsilon_{1}-1}{ \sqrt{ \sin^2\theta_{I}-1/\epsilon_{1} } } \; .
\end{equation}
%
%
%
In the Voigt configuration, we obtain
%
\begin{eqnarray}\label{DpVoigt}
D_{p}(\gamma) &=& \frac{\lambda \, \epsilon_{1}}{\pi} \frac{ e^{2\gamma} \, \sin\theta_{I} }{e^{4\gamma}\cos^2\theta_{I}-e^{2\gamma}\epsilon_{1}+\epsilon_{1}^2 \, \sin^2\theta_{I} } 
\nonumber \\
&&
\times\,\frac{1-e^{2\gamma}/\epsilon_{1}}{ \sqrt{ \sin^2\theta_{I}-e^{2\gamma}/\epsilon_{1} } } \; .
\end{eqnarray}

\begin{figure}[h]
\begin{centering}
\includegraphics[scale=0.6]{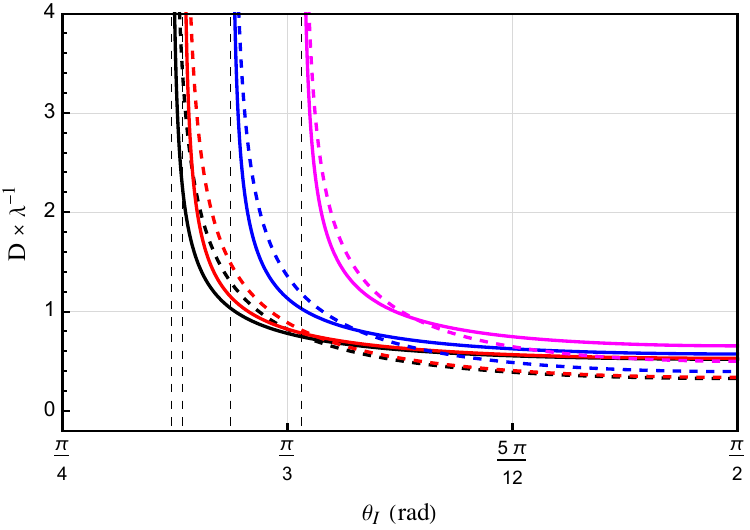}
\par\end{centering}
\caption{\small{\label{plot-GH-shift-ModMax} Goos-H\"anchen shift $D_{s}$ of \eqref{DsVoigt} for $s$-polarized incident wave (solid curves) and $D_{p}$ of \eqref{DpVoigt} (dashed lines) in terms of the incidence angle $\theta_{I}$. Here, we have used: $\epsilon_{1}=1.6$, $\gamma=0$ (usual scenario), $0.01$ (red), $0.05$ (blue), and $0.1$ (magenta). The vertical dashed lines indicate the critical angle, given by \eqref{critical-angle-GH-1},  above which the GH effect takes place.}}
\end{figure}

We illustrate the behavior of GH shift for $s$- and $p$-polarized incident waves in Fig.~\ref{plot-GH-shift-ModMax} as a function of incidence angle for configurations that depend on the $\gamma$ parameter.

One observes that, in all cases illustrated, the GH effect assumes a large value for incidence angles greater than the critical angle of total reflection $\theta_{c}$, which is given by
\begin{align}
\theta_{c}&=\arcsin\left[ \frac{e^{\gamma}}{\sqrt{\epsilon_{1}}} \right], \label{critical-angle-GH-1}
\end{align}
and then decreases monotonically toward its asymptotic value as the incidence angle approaches $\pi/2$. The standard scenario with $\gamma=0$ is represented by the black lines in Fig.~\ref{plot-GH-shift-ModMax} for $s$-polarized incident wave (solid) and $p-$polarized incident wave (dashed).

Furthermore, for a fixed incidence angle, we observe that increasing the magnitude of the non-linear parameter $\gamma$ yields an enhancement of the GH shift. In addition, the GH shift is slightly greater for $p$-polarized incident waves when the incidence angle is close to the critical angle $\theta_{c}$.

It is important to mention that, since the GH effect takes place when total reflection occurs, one may establish a condition on the possible values of $\gamma$ for which the GH effect can occur. In this sense, from $\sin \theta_{I} > \sin ( e^{\gamma} / \sqrt{\epsilon_{1}})$ one finds
\begin{align}
\gamma < \frac{1}{2} \ln (\epsilon_{1} \sin^{2}\theta_{I} ) , \label{GH-condition-1}
\end{align}
and considering that the maximum value of $\sin \theta_{I}$ is unity, the $\gamma$ parameter magnitude is constrained by
\begin{align}
\gamma < \frac{1}{2} \ln (\epsilon_{1}) . \label{GH-condition-2}
\end{align}
The latter represents a condition on the magnitude of $\gamma$ for which GH can take place.

In general, Figure \ref{plot-GH-shift-ModMax} indicates that the GH effect is shifted from the usual cases (black lines) for both $s-$ and $p-$polarized incident waves. On the other hand, increasing the magnitude of $\gamma$ parameter reduces the angular interval $\theta_{c} < \theta < \pi/2$ within which the effect is expected to occur.

\section{Complex Kerr rotation}
\label{sec5}
In this section, we consider the same situation: the interface between a linear dielectric medium, with electric permittivity $\epsilon_{1}$ and magnetic permeability $\mu_{1}$, and the second medium ruled by the permittivity and permeability tensors of ModMax electrodynamics. We can now discuss the polarization aspects regarding the reflected wave scattered off from the surface of the exotic medium, whose nonlinear constitutive relations are given by (\ref{eij})-(\ref{muijinv}) in the case in which the magnetic (B) background field is perpendicular to the electric (E) background. We will discuss the cases of $B > E$ and $E > B$.

\subsection{The case $B>E$}
Initially, we choose the condition of $B > E$, where the coefficients of the expansion around the EM background are reduced to  
%
%

%
\begin{subequations}
\label{modmax-coefficientes-EB-perpendicular-1}
\begin{align}
c_{1} & = e^{-\gamma} \; \; , \; \; d_{2}= \frac{2\,\sinh \gamma}{B^{2} - E^{2}} \; , 
\\
c_{2}&=d_{1}=d_{3}=0 \; , 
\\
d_{B} & = 0 \; \; , \; \; d_{E} = \frac{e^{2\gamma}-1}{B^{2} - E^{2}} \; .
\end{align}
\end{subequations}
In the case of incident light composed of $s$- and $p$-polarization states, $e_{\perp}$ and $e_{\parallel}$, respectively, we can use the general results of complex Kerr rotation given in the Ref. \cite{Pedro-Kerr-rotation}, which allows us to write the reflection matrix equation as
\begin{align}
\begin{pmatrix}
e_{R}^{\perp} \\
\\
e_{R}^{\parallel}
\end{pmatrix} 
&= \frac{1}{\Delta} \begin{pmatrix} 
r_{ss} && r_{sp} \\
\\
r_{ps} && r_{pp}
\end{pmatrix}
\begin{pmatrix}
e_{\perp} \\
\\
e_{\parallel} 
\end{pmatrix} \; , \label{general-reflection-matrix-1}
\end{align}
where $e_{R}^{\perp}$ and $e_{R}^{\parallel}$ are the $s$- and $p$-components of the reflected wave polarization. 
We choose, for simplicity, the fields as ${\bf{B}}= B \, \hat{\bf{x}}$ and ${\bf{E}}=E \, \hat{\bf{z}}$, the wave propagation direction as ${\bf n}_{2}=n_{2} \, (\cos\theta, 0, \sin \theta)$ in the medium (2). Then implementing these conditions and the constitutive relations (\ref{eij})-(\ref{muijinv}) in the reflection matrix (\ref{general-reflection-matrix-1}), we obtain
\begin{widetext}
\begin{align}
\begin{pmatrix}
e_{R}^{\perp} \\
\\
e_{R}^{\parallel}
\end{pmatrix} &= 
\begin{pmatrix}
 \displaystyle{  \frac{\sqrt{\mu_{1} \, \epsilon_{1}} \, \cos\theta_{I}- \mu_{1} \, n_{2} \, \cos\theta_{T}   }{  \sqrt{\mu_{1} \, \epsilon_{1}} \, \cos\theta_{I}+ \mu_{1} \, n_{2} \, \cos\theta_{T} }   }&& \displaystyle{\frac{ 2 \,  \mu_{1} \, \sqrt{\mu_{1}\epsilon_{1}} \, d_{E} \, E \, B \, \cos\theta_{I} \sin \theta_{I}}{\Delta} }\\
\\
0 &&   \displaystyle{ \frac{ \mu_{1} \, n_{2} \, \cos\theta_{I} - \sqrt{\mu_{1} \, \epsilon_{1}} \, \cos\theta_{T} }{\mu_{1} \, n_{2} \, \cos\theta_{I}
+\sqrt{\mu_{1} \, \epsilon_{1}}  \, \cos\theta_{T} } }
\end{pmatrix} 
\begin{pmatrix}
e_{\perp} \\
\\
e_{\parallel}
\end{pmatrix} \; , \label{reflection-matrix-modmax-new-1}
\end{align}
with
\begin{align}
\Delta &=  \left(\mu_{1} \, n_{2} \, \cos\theta_{T} + \sqrt{\mu_{1}\epsilon_{1}}  \cos \theta_{I} \right) \, \left( \mu_{1} \, n_{2} \, \cos\theta_{I} + \sqrt{\mu_{1}\,\epsilon_{1}} \, \cos\theta_{T} \right) \; , \label{reflection-matrix-modmax-new-2}
\end{align}
\end{widetext}
in which $\theta_{T}$ is the refracted angle, and $n_{2}$ is the refractive index solution (\ref{nomegap}) with sign $(+)$ 
\begin{eqnarray}
n_{2}^2= \frac{ B^{2} \, e^{2\gamma} - E^{2}}{(B^{2} - E^{2}) \left( \, \sin^2\theta + e^{2\gamma} \cos^2\theta \, \right)} \; , \label{index-nonlinear-with-E-B-1}
\end{eqnarray}
which contains the dependence with the ModMax $\gamma$-parameter, and of the $\theta$-angle. For an incident wave $s$-polarized, the complex Kerr rotation is described by the Kerr rotation angle $\theta_{K}$, and the Kerr ellipticity angle $\eta_{K}$, given by
\begin{align}
\tan (2\theta_{K}^{s}) &= \frac{\mathrm{Re}[2S]}{1- |S|^{2}} 
\hspace{0.3cm} \mbox{and} \hspace{0.3cm}
\sin (2\eta_{K}^{s}) = \frac{\mathrm{Im}[2S]}{1+ |S|^{2}} \; , \label{kerr-modmax-1}
\end{align}
where 
\begin{align}
S&= \frac{r_{ps}}{r_{ss}} \; . \label{Kerr-parameter-wave-type-S-1}
\end{align}
For the case of orthogonal electric and magnetic background fields (with $B>E$) in ModMax electrodynamics, the reflection matrix in \eqref{reflection-matrix-modmax-new-1} reveals that $S=0$, meaning that $\theta_{K}^{s}=0$ and $\eta_{K}^{s}=0$. The latter indicates that the reflected wave is linearly polarized, while the former indicates that polarization rotation does not occur. 
On the other hand, when the incident wave is $p$-polarized, the Kerr rotation angle, $\theta_{K}^{p}$, and Kerr ellipticity angle, $\eta_{K}^{p}$, are given by 
\begin{align}
\tan (2\theta_{K}^{p}) &= \frac{\mathrm{Re}[2P]}{1- |P|^{2}} 
\hspace{0.3cm} \mbox{and} \hspace{0.3cm} 
\sin (2\eta_{K}^{p}) = \frac{\mathrm{Im}[2P]}{1+ |P|^{2}} \; , \label{kerr-modmax-2}
\end{align}
where 
\begin{align}
P&= \frac{r_{sp}}{r_{pp}} \; , \label{Kerr-parameter-wave-type-P-1}
\end{align}
which reads
%
%
%
%
\begin{eqnarray}
P= \frac{2 \, \sqrt{\frac{\epsilon_{1}}{\mu_{1}}} \, d_{E} \, E \, B \, \tan \theta_{T}}{ n_{2}^{2} - \frac{\epsilon_{1}}{\mu_{1}} + n_{2} \, \sqrt{\frac{\epsilon_{1}}{\mu_{1}} } \, \chi } \; , \label{Kerr-parameter-wave-type-P-4} 
\end{eqnarray}
with
\begin{eqnarray}
\chi = \frac{\cos\theta_{I}}{\cos\theta_{T}} - \frac{\cos\theta_{T}}{\cos\theta_{I}} \; . \label{Kerr-parameter-wave-type-P-5}
\end{eqnarray}
The latter indicates that when incident light is $p$-polarized, the complex Kerr rotation takes place in the system. Notice that this effect takes place only when there are non-null electric and magnetic fields, which yields the mixed element $r_{sp}$ in the reflection matrix of (\ref{reflection-matrix-modmax-new-1}).

Considering the set chosen for external electromagnetic fields and propagation direction in medium 2 (nonlinear system ruled by ModMax electrodynamics), one can rewrite the $\theta$ angle in terms of transmission angle $\theta_{T}$ as $\theta=\pi/2 - \theta_{T}$. Afterwards, using Snell's law and the refractive index $n_{2}$ of \eqref{index-nonlinear-with-E-B-1}, the transmission angle can be cast as
\begin{subequations}
\label{transmission-angle-0}
\begin{align}
\theta_{T} &= \arcsin \left[ \, \sqrt{  \frac{ \mu_{1}\epsilon_{1}\sin^{2}\theta_{I} (B^{2} -E^{2})}{B^{2} e^{2\gamma} -E^{2} +g\, (B^{2} -E^{2})  \mu_{1}\epsilon_{1}\sin^{2}\theta_{I}}} \, \right] \; , \label{transmission-angle-1} 
\end{align}
with
\begin{align}
g = 1-e^{2\gamma} \; . \label{transmission-angle-2}
\end{align}
\end{subequations}

To illustrate the characteristic Kerr rotation and Kerr ellipticity angles of \eqref{kerr-modmax-2}, we implement \eqref{transmission-angle-0} and \eqref{index-nonlinear-with-E-B-1} in \eqref{Kerr-parameter-wave-type-P-4}, which yields
\begin{align}
\left. P \right|_{B>E} &= \frac{-2 \, y \, (e^{2\gamma}-1) \, \sqrt{ \frac{\epsilon_{1}}{\mu_{1}} } \, Q_{1} }{(y^{2}-1) \left[ \frac{ (e^{2\gamma}-1) \mu_{1} \epsilon_{1} (y^{2}-1) \sin^{2}\theta_{I} - y^{2} e^{2\gamma} +1}{y^{2}-1} +Q_{2} \right]} \; , \label{P-simplificado-1}
\end{align}
with
\begin{subequations}
\begin{eqnarray}
Q_{1} &=&  \sqrt{ \frac{\mu _1 \left(y^2-1\right) \epsilon _1 \sin ^2 \theta_{I} }{e^{2 \gamma } \mu _1 \left(1-y^2\right) \epsilon_1 \sin ^2 \theta_{I} +e^{2 \gamma } y^2-1} } \; , 
\\
Q_{2}  &=& \frac{\epsilon_{1}}{\mu_{1}} + \frac{ 1-y^{2}e^{2\gamma} - \mu_{1}\epsilon_{1} g (y^{2}-1) \sin^{2}\theta_{I}}{y^{2}-1} + Q_{3} \; , \\
Q_{3}  &=& \frac{ \sin \theta_{I} \sqrt{\frac{\epsilon_{1}}{\mu_{1}}} \, Q_{6} \, \tan \theta_{I} \, Q_{4}}{2(y^{2}-1) \, Q_{5} \, \sqrt{ e^{2\gamma}y^{2}-1} } \; , \\
Q_{4}  &=& \sqrt{ \frac{ (y^{2}-1) \, (e^{2\gamma}y^{2}-1)} {e^{2\gamma}y^{2}-1 +g \mu_{1}\epsilon_{1} (y^{2}-1) \sin^{2} \theta_{I} }} \; , \\
Q_{5}  &=& \sqrt{ \frac{ 1-e^{2\gamma}y^{2}+ e^{2\gamma} \mu_{1}\epsilon_{1} (y^{2}-1) \sin^{2}\theta_{I}}{1-e^{2\gamma}y^{2} -g \, \mu_{1}\epsilon_{1} (y^{2}-1) \sin^{2}\theta_{I} } } \; , \\
Q_{6}  &=& -2(1-e^{2\gamma}y^{2})-\mu_{1}\epsilon_{1} \,(y^{2}-1) \, [1+e^{2\gamma}+g \cos (2\theta_{I}) ]\; ,
\nonumber \\
\end{eqnarray}
\end{subequations}
where we have defined $y=B/E$.

\begin{figure}[h]
\begin{centering}
\includegraphics[scale=0.65]{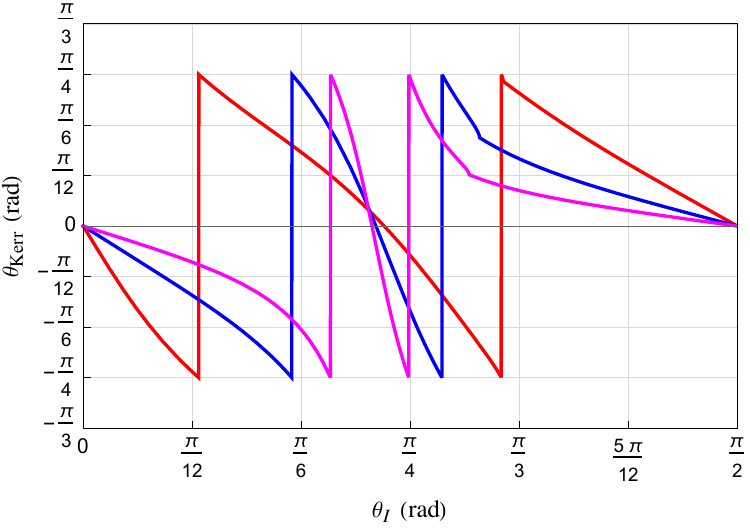}
\par\end{centering}
\caption{\small{\label{plot-Kerr-rotation-angle-1}}Kerr rotation angle $\theta_{K}$ of \eqref{kerr-modmax-2} in terms of the incidence angle $\theta_{I}$. Here, we have used: $\gamma=0.1$, $\mu_{1}=1$, $\epsilon_{1}=1.6$, and $y=$$1.5$ (red), $2$ (blue), and $3$ (magenta).}
\end{figure}

In Fig.~\ref{plot-Kerr-rotation-angle-1}, we depict the Kerr rotation angle $\theta_{K}$ of \eqref{kerr-modmax-2} in terms of the incidence angle $\theta_{I}$ for some values of the dimensionless parameter $y=B/E$.

We note that in each example there are two abrupt changes in the sign of $\theta_{K}$. This happens when $1-|P|^{2} \rightarrow 0$, which yields a divergence in the expression of $\theta_{K}$. This sign change (or reversion) indicates, in general, that the major axis of the elliptically polarized reflected wave will change its rotation direction since  $\theta_{K} >0 (<0)$ means counterclockwise (clockwise) direction rotation \cite{Hecht, Zangwill}. A similar behavior is also expected to occur in Weyl semimetals \cite{Sonowal}, whose electromagnetic response can be described within the axion electrodynamics framework. However, since the latter systems are dispersive, their abrupt sign change occur for specific frequencies, whereas for the nonlinear medium described in this work the sign change occurs at specific values of incidence angles $\bar{\theta}_{I}$ which are obtained by solving the equation $1-|P|^{2}=0$ for $\theta_{I}$ in each particular case. Furthermore, as the ratio $y=B/E$ increases, the values of $\bar{\theta}_{I}$ in which the abrupt sign change occurs progressively approach each other.

\begin{figure}[h]
\begin{centering}
\includegraphics[scale=0.65]{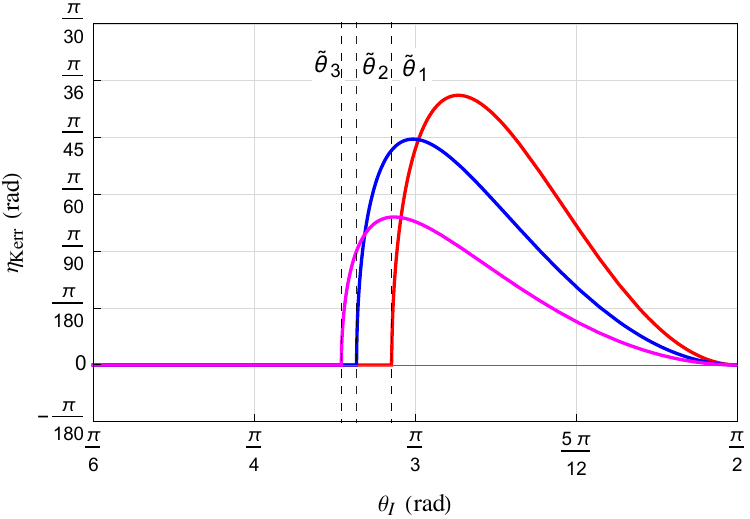}
\par\end{centering}
\caption{\small{\label{plot-Kerr-ellipticity-angle-1}}Kerr ellipticity angle $\eta_{K}$ of \eqref{kerr-modmax-2} in terms of the incidence angle $\theta_{I}$. Here, we have used: $\gamma=0.1$, $\mu_{1}=1$, $\epsilon_{1}=1.6$, and $y=$$1.5$ (red), $2$ (blue), and $3$ (magenta). The vertical dashed lines are given by $\tilde{\theta}_{i}$ of \eqref{expression-tilde-theta-1} for each example $i=$ 1 (red), 2 (blue), 3 (magenta).}
\end{figure}

The general behavior of Kerr ellipticity angle $\eta_{K}$ of \eqref{kerr-modmax-2} is depicted in Fig.~\ref{plot-Kerr-ellipticity-angle-1} in terms of the incidence angle $\theta_{I}$ for some values of $y=B/E$.

The non-null values of $\eta_{K}$ occur for $\theta_{I} > \tilde{\theta}_{i}$, where $\tilde{\theta}_{i}$ is given by
\begin{align}
\tilde{\theta}_{i}&= \arcsin \left[ \, \sqrt{ \frac{y^{2}-e^{2\gamma}}{\mu_{1} \epsilon_{1} (y^{2}-1) }} \, \right] \; , \label{expression-tilde-theta-1}
\end{align}
considering $\gamma>0$, $\mu_{1}>0$ and $\epsilon_{1}>0$. For the particular case with $\mu_{1}=\epsilon_{1}=1$, one finds $\eta_{K}=0$. The non-null values of $\eta_{K}$ occurring for $\theta_{I} > \tilde{\theta}_{i}$ indicate that the reflected wave will be left-handed elliptically \footnote{We are considering standard sign convention \cite{Zangwill, Jackson, Hecht}.} polarized ($\eta_{K} >0$). For $\theta_{I} < \tilde{\theta}_{i}$, the reflected wave is linearly polarized.

One observes that, as $y=B/E$ increases, the peak value of $\eta_{K}$ decreases. Moreover, the value of $\tilde{\theta}_{i}$ also decreases, indicating that Kerr ellipticity arises over a wider range of incidence angles, which may be referred to as the \textit{elliptical incidence window} $ \theta_{I} > \tilde{\theta}_{i} $, meaning a range of incidence angles for which the reflected wave becomes elliptically polarized.

It is worth mentioning that the magnitude of Kerr rotations is typically small, being less than 1 degree in usual materials \cite{Schlenker-Souche} and of the order $10^{-6}$ to $10^{-4}$ radians in topological insulators \cite{Sonowal}. Thus, the values obtained here can be regarded as {\it giant} in comparison with those material systems.

Another feature one notices is that, in Fig.~\ref{plot-Kerr-rotation-angle-1}, there is an incidence angle $\bar{\bar{\theta}}_{I}$, defined in $\bar{\theta}_{I(1)} < \bar{\bar{\theta}}_{I(1)} < \bar{\theta}_{I}$ (with $\theta_{I(1)}$ and $\theta_{I(2)}$ being solutions of $1-|P|^{2}=0$) where occurs a smooth sign reversion of Kerr rotation angle. One may interpret this sign change as a reversion in the polarization rotation of a linearly polarized reflected wave when the Kerr ellipticity angle $\eta_{K}$ is null. This means that, when $\eta_{K}=0$, the reflected wave will be rotated to the left or to the right depending on the value of incidence angle $\theta_{I}$, which can be found in two scenarios: $i)$ $\bar{\theta}_{I(1)} < \theta_{I} < \bar{\bar{\theta}}_{I}$, a linearly polarized reflected wave rotated in the counterclockwise direction (left); and $ii)$ $\bar{\bar{\theta}}_{I} < \theta_{I} < \bar{\theta}_{I(2)}$, the reflected wave will be rotated in the clockwise direction (right). Although an analytical expression for $\bar{\bar{\theta}}_{I}$ seems difficult to be obtained since the expression of $P$ in \eqref{P-simplificado-1} is very complicated and intricate, one can obtain the value of $\bar{\bar{\theta}}_{I}$, for each example in Fig.~\ref{plot-Kerr-rotation-angle-1}, by solving the equation
\begin{align}
-g \,\mu_{1}\, \epsilon_{1} (y^{2}-1) \,\sin^{2}\theta_{I} - y^{2} \,e^{2\gamma} +1 + (y^{2}-1)\, Q_{2} =0 \; ,
\end{align}
whose solution is
\begin{eqnarray}
\theta_{I}= \arcsin\left[ \, \sqrt{ \frac{1-Q_{2}+y^2\,(Q_2-e^{2\gamma})}{g\,\mu_1\,\epsilon_1\,(y^2-1)} } \, \right] \; .
\end{eqnarray}
\subsection{The case $E > B$}
\begin{figure}[h]
\begin{centering}
\includegraphics[scale=0.65]{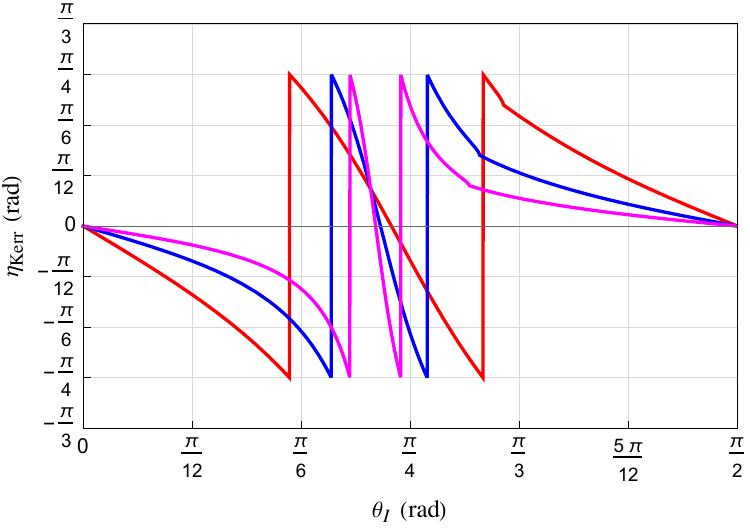}
\par\end{centering}
\caption{\small{\label{plot-Kerr-rotation-E-maior-B}Kerr rotation angle $\theta_{K}$ of \eqref{kerr-modmax-2} in terms of the incidence angle $\theta_{I}$. Here, we have used: $\gamma=0.1$, $\mu_{1}=1$, $\epsilon_{1}=1.6$, and $x=$$1.5$ (red), $2$ (blue), and $3$ (magenta).}}
\end{figure}

In this case of $E > B$, the coefficients of the expansion are given by
\begin{subequations}
\label{modmax-coefficientes-EB-perpendicular-2}
\begin{align}
c_{1} & = e^{\gamma} \; \; , \; \;  
\\
c_{2}&=d_{1}=d_{3}=0 \; , 
\\
d_{B} & = 0 \; \; , \; \; d_{E} = \frac{1-e^{-2\gamma}}{E^{2} - B^{2}} \; ,
\end{align}
\end{subequations}
in which the refractive index solution for ${\bf{B}}= B \, \hat{\bf{x}}$, ${\bf{E}}=E \, \hat{\bf{z}}$, and ${\bf n}_{2}=n_{2} \, (\cos\theta, 0, \sin \theta)$ 
in the medium (2) is
%
%
%
\begin{align}
n_{2}^{2} &= \frac{E^{2}e^{2\gamma} - B^{2}}{(E^{2}-B^{2}) (e^{2\gamma}\sin^{2}\theta + \cos^{2}\theta )} \; . \label{index-refraction-E-maior-que-B-1}
\end{align} 
Analogously to the procedure of the previous section, one finds now the refractive angle
\begin{align}
\theta_{T} &= \arcsin \left[ \, \sqrt{ \frac{ \mu_{1}\epsilon_{1} \sin^{2}\theta_{I} (E^{2}-B^{2})}{E^{2}\,e^{2\gamma}-B^{2} - g \, (E^{2} -B^{2}) \, \mu_{1}\epsilon_{1} \sin^{2}\theta_{I}}} \, \right] \; , \label{refraction-angles-E-maior-que-B-1}
\end{align}
with $g$ defined in \eqref{transmission-angle-2}.
Implementing \eqref{refraction-angles-E-maior-que-B-1} and \eqref{index-refraction-E-maior-que-B-1} in \eqref{Kerr-parameter-wave-type-P-4}, we obtain
\begin{align}
\left. P \right|_{E>B} &=\frac{ -2 \, (e^{-2\gamma}-1) \, x \, \sqrt{ \frac{\epsilon_{1}}{\mu_{1}}} \, U_{1} }{ (x^{2}-1)  \left(U_{2} + U_{6} \right) } \; , 
\label{P-case-E-maior-B-1}
\end{align}
where
\begin{subequations}
\begin{eqnarray}
U_{1} &=& \sqrt{ \frac{ \mu_{1}\epsilon_{1} (x^{2}-1) \sin^{2}\theta_{I}}{x^{2}e^{2\gamma}-1-\mu_{1}\epsilon_{1}e^{2\gamma}(x^{2}-1) \sin^{2}\theta_{I}}} \; ,  
\\
U_{2} &=& -\frac{\epsilon_{1}}{\mu_{1}} + U_{3} \; , 
\\
U_{3} &=& \frac{ 2 (-1+x^{2}\, e^{2\gamma}) \, U_{4} } {(y^{2}-1) \, (2e^{2\gamma} -2 e^{4\gamma}x^{2} +U_{5} ) } \; , 
\\
U_{4} &=&  1-x^{2} e^{2\gamma} - g\mu_{1}\epsilon_{1} (x^{2}-1) \sin^{2}\theta_{I} \; , 
\\
U_{5} &=& (-1-e^{2\gamma} + 2e^{4\gamma}) \, \mu_{1}\epsilon_{1} (x^{2}-1) \sin^{2}\theta_{I} \; , 
\\
U_{6} &=& \frac{ \sqrt{\frac{\epsilon_{1}}{2\mu_{1}}} \sin\theta_{I} \tan \theta_{I} \sqrt{ x^{2} \, e^{2\gamma}-1} \, U_{7} \, U_{8}} {[1-x^{2}e^{2\gamma} + e^{2\gamma} \mu_{1}\epsilon_{1} (x^{2}-1) \sin^{2}\theta_{I} ]U_{9} } \; , 
\\
U_{7} &=& -2+2x^{2}e^{2\gamma}-(x^{2}-1) \mu_{1}\epsilon_{1}[1+e^{2\gamma} +g \cos(2\theta_{I}) ] \; ,
\nonumber \\
\\
U_{8} &=& \sqrt{ \frac{1-x^{2} e^{2\gamma} +e^{2\gamma} \mu_{1}\epsilon_{1} (x^{2}-1) \sin^{2}\theta_{I}}{ 1-x^{2}e^{2\gamma} - g \mu_{1}\epsilon_{1} (x^{2}-1) \sin^{2}\theta_{I}}} \; , 
\\
U_{9} &=& \sqrt{ \frac{ (x^{2}-1) (2e^{2\gamma} - 2x^{2} \, e^{4\gamma} + U_{10} )}{1-x^{2} \, e^{2\gamma} - g \, \mu_{1}\epsilon_{1} (x^{2}-1) \sin^{2}\theta_{I}}} \; , 
\\
U_{10} &=&  \mu_{1}\epsilon_{1} (x^{2}-1) (-1-e^{2\gamma} + 2e^{4\gamma} )\sin^{2}\theta_{I} \; .
\end{eqnarray}
\end{subequations}
Using now \eqref{P-case-E-maior-B-1}, we illustrate the general behavior of $\theta_{K}$ in terms of incidence angle $\theta_{I}$ in Fig.~\ref{plot-Kerr-rotation-E-maior-B} for some values of the dimensionless parameter $x=E/B$.

As before, one also notices that there are two abrupt sign changes for $\theta_{K}$, which occur at incidence angles $\bar{\theta}_{I}$ that are solutions of $1-|P|^{2}=0$ with $P$ of \eqref{P-case-E-maior-B-1}. Moreover, as $x=E/B$ increases, the two divergences approach each other progressively. On the other hand, for $E>B$, these two angles are closer to each other when compared to the case of $B >E$ for the same set of electromagnetic parameters.

\begin{figure}[h]
\begin{centering}
\includegraphics[scale=0.65]{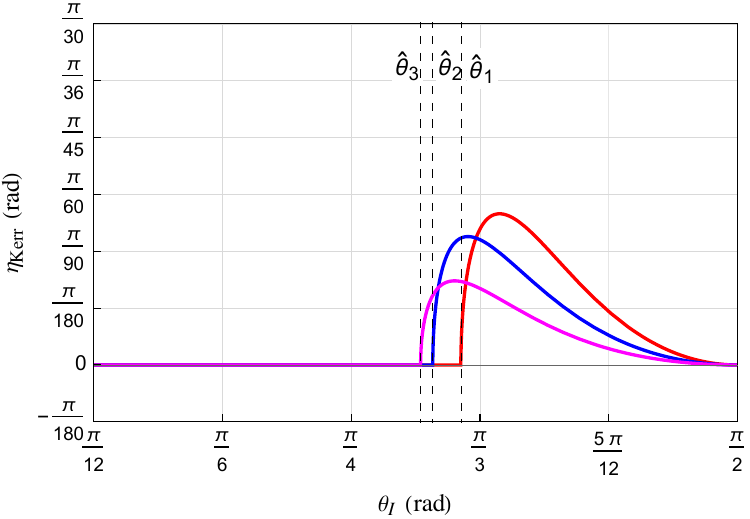}
\par\end{centering}
\caption{\small{\label{plot-Kerr-ellipticity-E-maior-B}Kerr ellipticity angle $\eta_{K}$ of \eqref{kerr-modmax-2} in terms of the incidence angle $\theta_{I}$. Here, we have used: $\gamma=0.1$, $\mu_{1}=1$, $\epsilon_{1}=1.6$, and $x=$$1.5$ (red), $2$ (blue), and $3$ (magenta). The vertical dashed lines indicate the incidence angle $\hat{\theta}_{i}$ of \eqref{incidence-kerr-ellipticity-E-maior-B-1}.}}
\end{figure}

Regarding the Kerr ellipticity, for $E> B$ the peak of $\eta_{K}$ is smaller in comparison with $B >E$ scenario, as shown in Fig.~\ref{plot-Kerr-ellipticity-E-maior-B}. This means that the polarization ellipse of the reflected wave is more eccentric than that expected in the scenario of $B>E$. The elliptical incidence window in this case is defined by $\theta_{I} > \hat{\theta}_{i}$, with $\hat{\theta}_{i}$ given by
\begin{align}
\hat{\theta}_{i}&= \arcsin \left[ \, \sqrt{ \frac{x^{2} - e^{2\gamma}}{\mu_{1}\epsilon_{1} \, (x^{2}-1)}} \, \right] \; , \label{incidence-kerr-ellipticity-E-maior-B-1}
\end{align}
in which $i$-index $(i=1,2,3)$ symbols the colors red, blue, and magenta in the Fig. \ref{plot-Kerr-ellipticity-E-maior-B}.


%
\section{Final remarks}
\label{sec6}

In this work, we have investigated some optical properties arising in systems ruled by the modified Maxwell (ModMax) electrodynamics. First, we have obtained the general dispersion relations and modified refractive indices considering uniform external electric and magnetic backgrounds. In the case of a uniform magnetic background, the modified refractive index presents a $\theta$ dependence, which represents the angle between the direction of propagation and the magnetic background. In this scenario, the propagating modes are, generally, given by linearly polarized vectors. Thereby, the birefringence can be evaluated in terms of phase shift of \eqref{eq-nonlinear-17}, which has linear dependence on the non-linear ModMax $\gamma$-parameter, in the limit of small $\gamma$.
Regarding reflection properties, we have also investigated the Goos-H\"anchen (GH) shift in the scenario of a uniform magnetic background in Sec.~\ref{sec4}. For $s$-polarized incident wave, the GH effect is given, generally, by \eqref{Ds}, whereas for $p$-polarized incident wave, the GH shift is given by \eqref{Dp}.

To investigate the Kerr rotation, we have considered the case with perpendicular uniform electric (E) and magnetic (B) background fields with $B>E$ and $E>B$ in Sec.~\ref{sec5}. In the scenario of $B>E$, one finds that for $s$-polarized incident wave, the Kerr rotation angle ($\theta_{K}$) and Kerr ellipticity ($\eta_{K}$) are null, regardless of the magnitude of background fields and non-linear parameter $\gamma$. On the other hand, for $p$-polarized incident wave, Kerr rotation takes place for oblique incidence angles $\theta_{I}$ and $\gamma>0$, as indicated in \eqref{P-simplificado-1}. The abrupt sign change of $\theta_{K}$ happens for specific incidence angles $\bar{\theta}_{I}$ obtained by solving $1-|P|^{2}=0$, with $P$ given by \eqref{P-simplificado-1}. This abrupt sign change is also expected to occur in Weyl semimetals, whose electromagnetic response is described by the axion electrodynamics. In such a scenario, the axion terms yield the Kerr rotation. In the case investigated in this work, however, the Kerr signals occur due to a non-null uniform electromagnetic background and the non-linear parameter $\gamma$.

Furthermore, Kerr ellipticity angle $\eta_{K}$ assumes non-null values for $\theta_{I} > \tilde{\theta}_{i}$, with $\tilde{\theta}_{i}= \arcsin \left[ \sqrt{ \frac{y^{2}-e^{2\gamma}}{\mu_{1} \epsilon_{1} (y^{2}-1) }} \right] $, and $y=B/E$, which may be regarded as the elliptical incidence window, {\it i. e.}, the reflected wave becomes elliptically polarized when the incidence angle lies within the range $\theta_{I} > \tilde{\theta}_{i}$. This feature may work as a polarization conversion mechanism: $p$-polarized incident wave may be transformed into a left-handed elliptically polarized wave by reflecting on a non-usual medium ruled by the ModMax ED. In the particular cases where $\eta_{K}=0$, the reflected wave remains linearly polarized.

In the regime $E>B$, the Kerr rotation angle $\theta_{K}$ and ellipticity $\eta_{K}$ behave analogously to the case observed when $B >E$. However, the peak value of $\eta_{K}$  for $E >B$ is smaller than that of $B>E$, meaning that the polarization ellipse for the reflected wave is more eccentric in this case. The corresponding elliptical incidence window is now defined by $\theta_{I} > \hat{\theta}_{i}$, with $\hat{\theta}_{i}= \arcsin \left[ \sqrt{ \frac{x^{2}-e^{2\gamma}}{\mu_{1} \epsilon_{1} (x^{2}-1) }} \right] $, and $x=E/B$.

In general, our results show that the non-linear parameter $\gamma$, together with the quantities $y=B/E$ and $x=E/B$, play a key role in controlling the magnitude of Kerr signals (rotation and ellipticity), as well as the angular interval in which elliptically polarized light emerges. These properties suggest that ModMax electrodynamics can provide a consistent framework for investigating and describing non-usual material systems with optical effects induced by nonlinear electromagnetic effects.

Finally, we summarize some of the main properties obtained for non-usual medium ruled by ModMax electrodynamics in Tab.~\ref{tab:optical-effects-in-ModMax}.

\begin{widetext}

\begin{table}[H]
		\caption{Some optical effects in ModMax electrodynamics}
		\centering
		\begin{tabular}{p{4cm} p{12cm}}
				\hline \hline \\[-2ex]
				 & {\bf{Optical effect}} 
				 \\[0.5ex]
				 \hline
				 \\[-2ex]
				 Birefringence & yes. Refractive indices are given by \eqref{nomegap}. 
				 \\[.5ex]
				 \hline
				 \\[-2ex]
				 Propagation modes & Linearly polarized vectors for uniform magnetic background.
				  \\[.5ex]
				 \hline
				\\[-2ex]
				 Goos-H\"anchen shift &  yes. The GH effect is given by \eqref{Ds} and \eqref{Dp} for $s$- and $p$-polarized incident waves, respectively. For $p$-polarized incident waves, the GH shift is slightly enhanced compared to GH for $s$-polarized waves.
				 \\[.5ex]
				 \hline
				 \\[-2ex]
				 Complex Kerr rotation &  yes. It takes place for non-null electric and magnetic backgrounds and oblique incidence angles. For $s$-polarized incident wave, $\eta_{K}=0$ and $\eta_{K}=0$. For $p$-polarized incident wave, $\theta_{K}$ and $\eta_{K}$ are given by \eqref{kerr-modmax-2} with $P$ defined in two cases: i) $P$ of \eqref{P-simplificado-1} for $ B> E$; ii) $P$ of \eqref{P-case-E-maior-B-1} for $E >B$.			
				   \\[2ex]
\hline \hline
			\end{tabular}
		\label{tab:optical-effects-in-ModMax}
	\end{table}

\end{widetext}

\section*{Acknowledgements}
The authors express their gratitude to FAPEMA, CNPq, and CAPES (Brazilian research agencies) for their invaluable financial support. P.D.S.S. is grateful to grant Produtividade/FAPEMA/CNPq/12649/25. Furthermore, we are indebted to CAPES/Finance Code 001.

\end{document}